# A Microprocessor implemented in 65nm CMOS with Configurable and Bit-scalable Accelerator for Programmable In-memory Computing

Hongyang Jia, Yinqi Tang[1], Hossein Valavi[1], Jintao Zhang, Naveen Verma

*Princeton University, Princeton NJ*

***Abstract:*** This paper presents a programmable in-memory-computing processor, demonstrated in a 65nm CMOS technology. For data-centric workloads, such as deep neural networks, data movement often dominates when implemented with today's computing architectures. This has motivated spatial architectures, where the arrangement of data-storage and compute hardware is distributed and explicitly aligned to the computation dataflow, most notably for matrix-vector multiplication. In-memory computing is a spatial architecture where processing elements correspond to dense bit cells, providing local storage and compute, typically employing analog operation. Though this raises the potential for high energy efficiency and throughput, analog operation has significantly limited robustness, scale, and programmability. This paper describes a 590kb in-memory-computing accelerator integrated in a programmable processor architecture, by exploiting recent approaches to charge-domain in-memory computing [1]. The architecture takes the approach of tight coupling with an embedded CPU, through accelerator interfaces enabling integration in the standard processor memory space. Additionally, a near-memory-computing datapath both enables diverse computations locally, to address operations required across applications, and enables bit-precision scalability for matrix/input-vector elements, through a bit-parallel/bit-serial (BP/BS) scheme. Chip measurements show an energy efficiency of 152/297 1b-TOPS/W and throughput of 4.7/1.9 1b-TOPS (scaling linearly with the matrix/input-vector element precisions) at $V_{DD}$ of 1.2/0.85V. Neural network demonstrations with 1-b/4-b weights and activations for CIFAR-10 classification consume 5.3/105.2 µJ/image at 176/23 fps, with accuracy at the level of digital/software implementation (89.3/92.4 % accuracy).

## 1. INTRODUCTION

Neural-network inference is dominated by computation of high-dimensionality matrix-vector multiplications (MVMs). While hardware acceleration has typically enabled significant increase in energy efficiency and performance of compute, high-dimensionality of MVMs makes data movement a significant cost, limiting the gains achievable from traditional digital acceleration. For instance, embedded memory, often relied on to exploit opportunities for data reuse, can dominate energy and delay over actual compute operations for even modest sized arrays, due to the costs of data-movement from the point of storage to the point of compute outside the array. This has motivated spatial architectures (e.g., systolic arrays), where storage and computation hardware is distributed in processing elements (PEs) arranged in a 2D array, explicitly having structural alignment with the dataflow in MVMs. Both the bottlenecks imposed by memory and the benefits of such structural alignment have recently motivated architectures based on in-memory computing. In terms of memory bottlenecks, in-memory computing enables accessing of a computation result over many stored bits, rather than accessing of individual bits, thereby amortizing the accessing energy and delay. In terms of spatial architectures, in in-memory computing, bit cells correspond to highly energy- and area-efficient PEs, where the costs of accessing from local storage, performing compute, and moving data to the next element are significantly reduced compared to digital PEs.

    The primary challenge with in-memory computing is integrating compute in constrained memory circuits. This has required going beyond restrictive switched-based abstractions associated with digital compute to richer abstractions based on analog compute. However, analog compute, in previous designs, has limited the robustness, scale, and programmability achieved. Recent work [1] has moved from current-domain

---

[1] Equal contributing authors.

analog in-memory computing to charge-domain analog in-memory computing. While current-domain compute relies on transistor currents as the output signal from bit cells, charge-domain compute stores the output signal from bit cells as voltage on a localized capacitor. This directly corresponds to charge through the capacitance value (Q=C×V), and it has the benefit that capacitance can be precisely controlled in modern VLSI technologies (as it is primarily set by geometric parameters, thus benefitting from lithographic precision, in contrast to transistor parameters, which are subject to significant levels of semiconductor-device variations). Thus, in [1], moving to charge-domain compute substantially addressed robustness and scale. This work attempts to address programmability, by integrating charge-domain in-memory computing with near-memory interfaces, for configurability and integration in a microprocessor architecture, as well as near-memory computing, for bit-precision scalability and flexible localized post-reduce compute.

## 2. ARCHITECTURE AND CIRCUITS

Fig. 1 shows the programmable processor architecture, including: (1) a 590kb Compute-In-Memory Unit (CIMU), partitioned into 16 (4×4) banks, as well as configurable digital periphery and near-memory compute datapath; (2) 128kB of standard program/data memory (P/DMEM); (3) 2-channel DMA; (4) RISC-V CPU [2]; and (5) peripherals for external-memory control, bootloading, host-PC interfacing (UART), general-purpose IO (GPIO), and scheduling (timer).

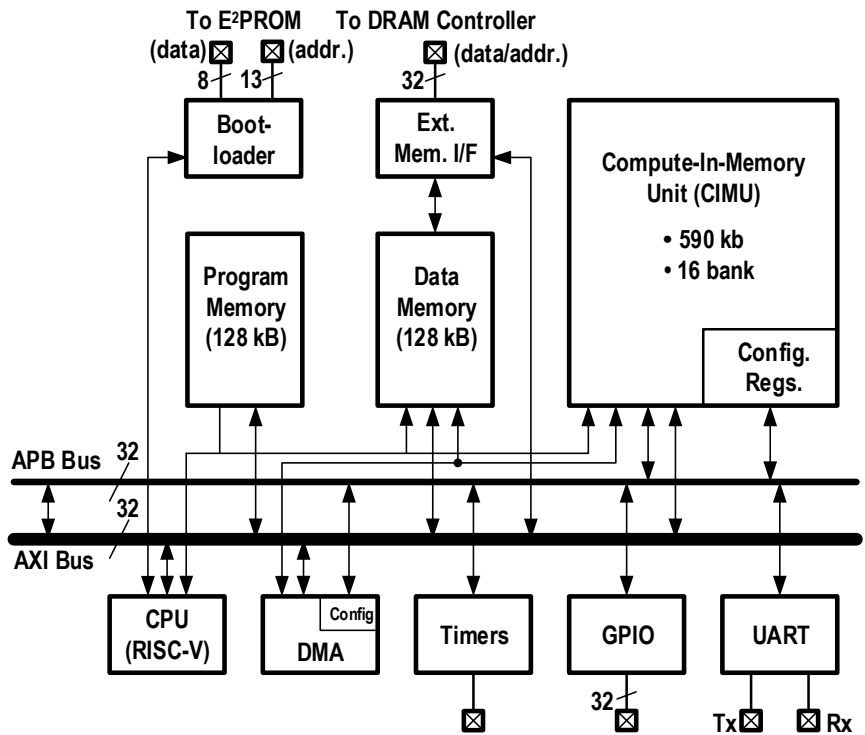

**Fig. 1. Architecture block diagram of programmable in-memory computing processor.**

The central block is the CIMU, shown in Fig. 2. Within this, the Compute-In-Memory Array (CIMA) performs mixed-signal matrix-vector multiplication ($\vec{y} = A\vec{x}$), having $\vec{x}$ dimensionality up to 3*3*256=2304 (especially to support 3×3 CNN filters) and $A$ dimensionality up to 256×2304, with both dimensionalities configurable via activity-gating of CIMA banks.

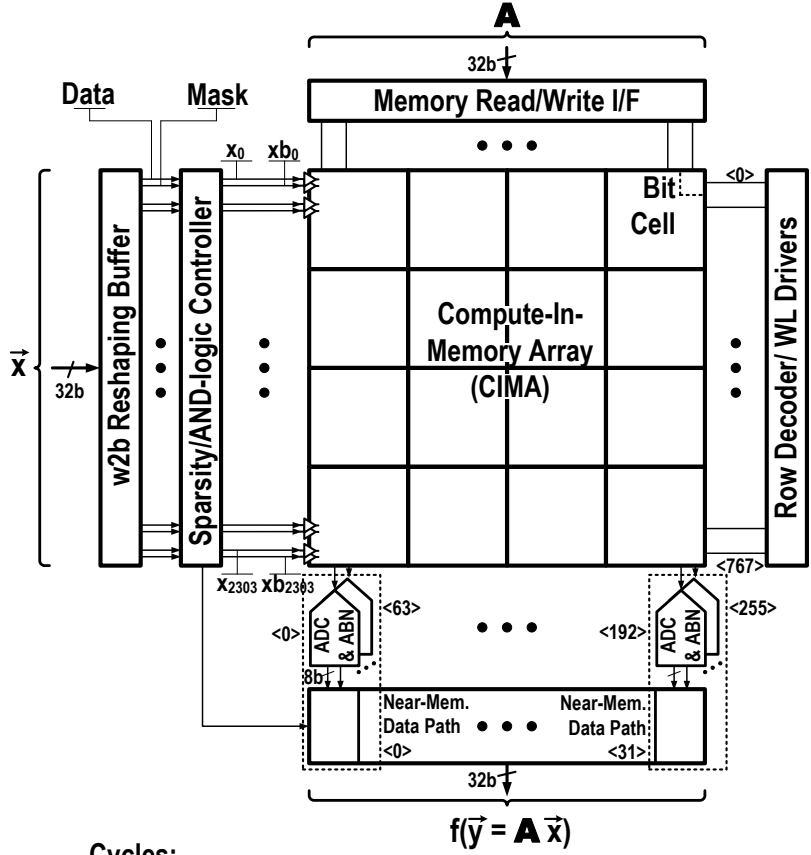

**Cycles:**
$C_{RD/WR}$= 20; $C_{CIMA}$= 50; $C_{ADC}$= 20; $C_{ABN}$= 20; $C_{NEAR-MEM}$=8

Fig. 2. Block diagram of Compute-In-Memory Unit (CIMU).

The CIMA is based on the charge-domain in-memory-computing approach from [1], employing the bit cell shown in Fig. 3, to implement multiplication and accumulation. In addition to a standard 6T SRAM cell for local storage, this includes two PMOS transistors and a capacitor for local multiplication, and CMOS (NMOS+PMOS) shorting switch for accumulation across a CIMA column. We describe the operation of one CIMA column, but all columns operate in parallel, computing $\vec{y} = A\vec{x}$ at once. Operation starts by shorting all local capacitors in the CIMA column together (by asserting $S/Sb$), and collectively discharging them to 0V (while holding $x_n/xb_n$ bits high, to disable the PMOS transistors). Then, after de-asserting $S/Sb$, local compute is performed, corresponding to binary multiplication, i.e., XNOR, between stored 1-b matrix element $a_{m,n}/ab_{m,n}$ bits and an inputted 1-b vector element $x_n/xb_n$ bits. The 1-b outputs $o_{m,n}$ are then stored as charge on the bit cells' local metal-oxide-metal (MOM) capacitor, laid out above the transistors (thus consuming no area). Finally, accumulation is performed by shorting together charge from all bit-cell capacitors in each CIMA column by asserting $S/Sb$.

Thus, this approach performs binary multiplication in the digital voltage domain, exploiting the high efficiency of transistor switching and inherent linearity of a 1-b output (i.e., two states imply no deviation from line). Yet, it performs high-dynamic-range accumulation in the analog charge domain, exploiting excellent process/temperature stability of capacitors, to eliminate switching costs incurred in a high-dynamic range digital accumulator (which would require >13 b in this design), thus efficiently amortizing the bit-by-bit accessing costs otherwise incurred in standard memory.

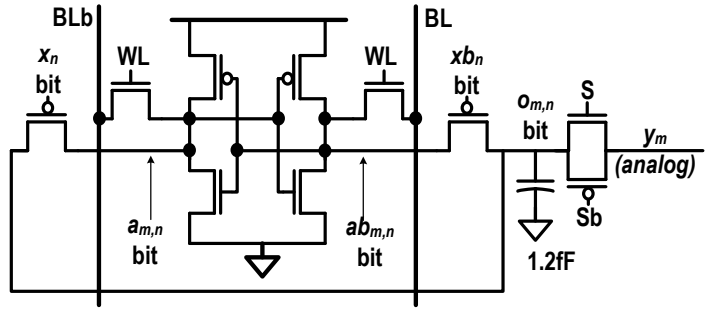

**Fig. 3. Charge-domain-computing bit cell [1].**

While the binary multiplications in [1] could efficiently support applications such as binarized neural networks (BNNs), for a programmable architecture meant to address broader applications, the CIMU extends to multi-bit compute. This is done via a bit-parallel/bit-serial (BP/BS) scheme, so that the efficient and linear bit-wise mixed-signal operation of CIMA columns can still be exploited. The BP/BS scheme is shown in Fig. 4, where $B_A$ corresponds to the number bits in the matrix elements, $B_X$ corresponds to the number of bits in the input-vector elements, and N corresponds to the dimensionality of the input vector ($M_n$ is a mask bit, used for sparsity and dimensionality control, as described later). The multiple bits of $a_{m,n}$ are mapped to parallel CIMA columns and the multiple bits of $x_n$ are inputted serially. Multi-bit multiplication and accumulation can then be achieved either via bit-wise XNOR or via bit-wise AND, both of which are supported by the bit cell of Fig. 2; specifically, bit-wise AND is achieved by driving only the $xb_n$ bit to the bit cell, and holding the $x_n$ bit high (see Fig. 3, not shown in Fig. 4 for simplicity). While bit-wise AND can support standard 2's complement number representation, bit-wise XNOR requires slight modification of the number representation (i.e., element bits map to +1/-1 rather than 1/0, necessitating two bits with LSB weighting to properly represent zero). Following each bit-wise CIMA-column operation, the output is then digitized and properly bit-shifted, before being added to other digitized and bit-shifted CIMA-column outputs, in order to derive the multi-bit operation. Thus, in the mixed-signal BP/BS scheme, the energy and area-normalized throughput scale linearly with the number of bits used for matrix and input-vector elements ($B_A \times B_X$); this is more efficient than the exponential scaling typically expected with purely analog schemes.

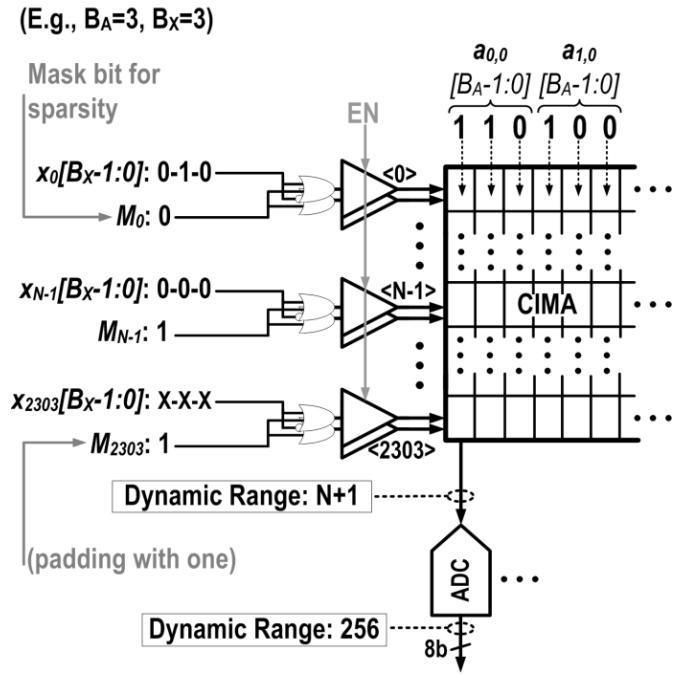

**Fig. 4. Bit-parallel/bit-serial (BP/BS) scheme for bit-scalable matrix and input-vector elements.**

Both to combine bit-wise CIMA-column operations into multi-bit operations, and to perform configurable localized compute required across broad applications, the CIMU employs the near-memory datapath shown in Fig. 5. Each CIMA column feeds an 8-b successive-approximation register (SAR) analog-to-digital converter (ADC) and a binarizing analog batch normalization (ABN) block (ABN is similar to [1], employing a 6-b DAC for analog reference generation). While the ABN supports BNNs, the ADC supports a range of subsequent post-reduce near-memory digital computation. Specifically, we note that while high-dimensionality MVMs necessitate in-memory computing to address data-movement costs, typically other computations are substantially addressed by traditional digital acceleration. The ADC resolution is selected to balance dynamic-range requirements (discussed below) and energy/area overheads, with the 8-b ADC introducing 18/15 % area/energy increase of a CIMA column. The digital datapath following the ADC/ABN is multiplexed across 8 CIMA columns and ADCs, as shown. This is done for area savings and throughput matching (cycle numbers shown in Fig. 2). The datapath provides: (1) BP/BS compute for any bit precision, via digital barrel shifting and summation of digitized CIMA-column outputs in time and space; and (2) other post-reduce compute, especially supporting neural-network acceleration (global/local scaling/biasing, batch normalization, activation function). Note, purely analog multi-bit charge-domain compute is also possible, but requires exponential-weighting of capacitors, degrading energy/area.

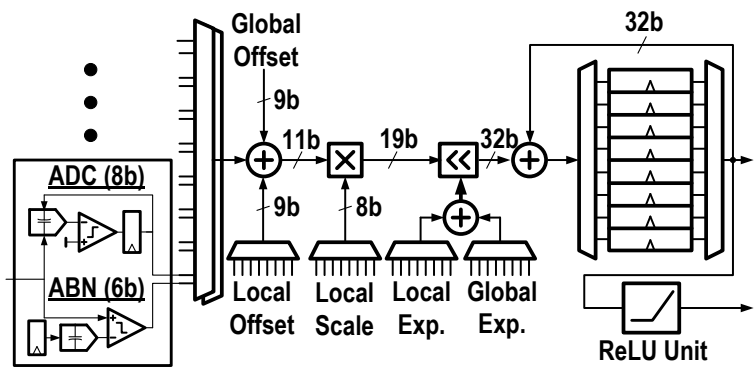

Fig. 5. Per-column ADC and ABN, and 8-way multiplexed near-memory digital datapath.

In addition to in/near-memory compute, the CIMU includes specialized interfaces for dataflow in a programmable architecture. While a number of approaches exist for accelerator integration with a CPU, here tight accelerator-CPU coupling is pursued to address programmability, where interfaces are included to integrate the CIMU within the standard processor memory space. First, for input-vector elements $x_n$, Fig. 6a shows the word-to-bit (w2b) Reshaping Buffer, which enables interfacing of external 32-b words to internal bit-wise operations. This minimizes data transfer to the CIMU by loading 1-to-8-b segments of incoming 32-b words into double-buffering register files, whose parallel readout then provides bit-serial inputting to the CIMA. For implementing convolutional neural networks (CNNs), such buffering also enables input-element shifting for striding, reducing input-activation loading to only new pixel locations. Second, Fig. 6b shows the Sparsity/AND-logic Controller, which masks $x_n/xb_n$ broadcasting over the CIMA array, both to exploit energy savings proportional to the number of zero-valued elements and to support bit-wise AND operations in the bit cells, as described above. While sparsity-proportional energy savings are inherently achieved with bit-wise AND operations (due to 2's compliment multi-bit representation), for bit-wise XNOR operations zero-valued elements are explicitly located, to derive a mask bit $M_n$ to prevent broadcasting of all bits, and tallied, to provide an offset to the near-memory datapath required in order to account for capacitors left in their reset state. In addition to reducing energy (i.e., preventing $x_n/xb_n$ broadcast and XNOR/AND compute energy, which account for ~50% of CIMA energy), exploiting sparsity in this way also benefits signal-to-quantization-noise ratio (SQNR), considered below. Third, for loading matrix-elements $a_{m,n}$, a local buffer interfaces external 32-b words with 768-b parallel CIMA writes.

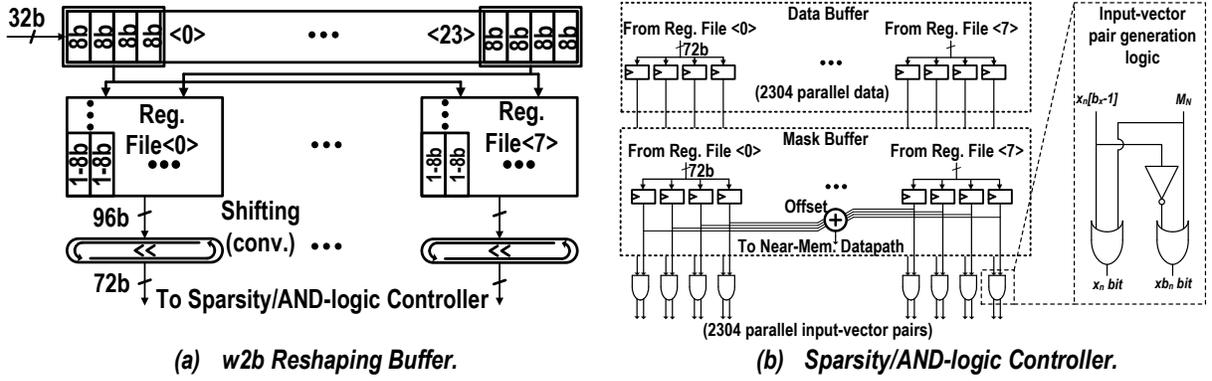

(a) w2b Reshaping Buffer.  (b) Sparsity/AND-logic Controller.

Fig. 6. CIMU interfaces for integration in standard processor memory space.

## 3. DESIGN ANALYSIS

Each CIMA column generates a high dynamic range analog signal, whose number of possible levels, set by the input-vector dimensionality N, is N+1 (i.e., accumulation of N different binary outputs from bit cells), with N being up to 2304 in this design. However, the 8-b ADC supports a dynamic range of up to 256 levels, chosen to limit its energy and area overhead. This mismatch impacts the SQNR of computation differently than standard integer computation. Specifically, Fig. 7a shows the simulated SQNR for bit-wise XNOR and Fig. 7b shows the simulated SQNR for bit-wise AND, both for different $B_X$, as $B_A$ is scaled (SQNR's for XNOR/AND are different due to dynamic range of number representation formats). If N is explicitly limited to 255, through CIMA-bank activity gating, or the number of CIMA-column output levels is implicitly limited to 255, through sparsity control, then the ADC dynamic range enables integer compute to be perfectly emulated, as shown. However, in all other cases, the SQNR is set not only by $B_A/B_X$, but also by N and the level of sparsity. Nonetheless, we point out that with 8-b ADC resolution, SQNR near standard integer compute is observed at operand quantization of typical interest in neural networks (2-6b).

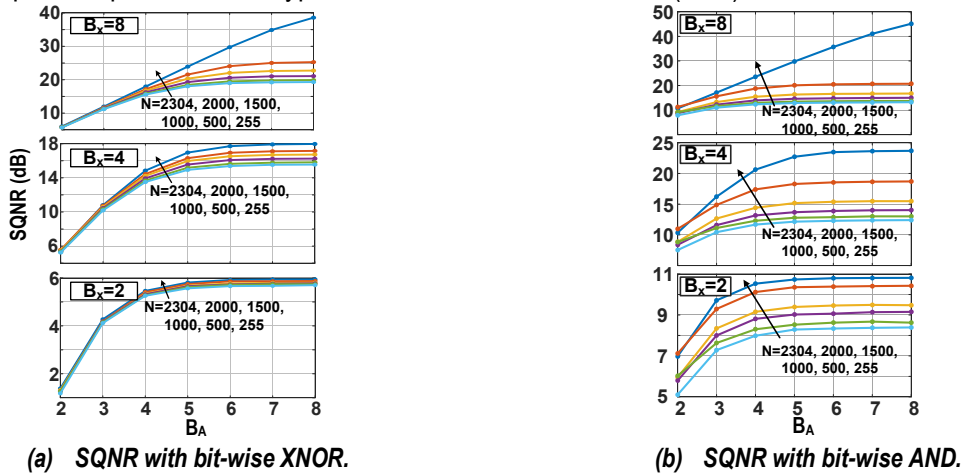

(a) SQNR with bit-wise XNOR.  (b) SQNR with bit-wise AND.

Fig. 7. SQNR analysis with respect to $B_A$, $B_X$, and N.

Fig. 8 analyzes data bandwidth, affecting utilization, to/from the CIMU via 32-b DMA transfers (taking ~1 cycle). First, considering $\vec{x}/\vec{y}$, the transfer cycles $C_x/C_y$ depend on the bit precisions $B_x/B_y$ and vector dimensionalities N/M. $C_x/C_y$ are shown along with the CIMU cycles $C_{CIMU}$ for different bit precisions (the near-memory datapath sets $B_y$ to 16 b if $B_X+B_A \leq 5$, else 32 b), at the maximum vector dimensionalities N=2304 and M=256/$B_A$ ($B_A$ is number of matrix-element bits). As seen, $C_{CIMU}$ is typically highest, giving high CIMU utilization by pipelining data transfers and CIMU operation; however, considerable potential to push CIMU speed may eventually make dedicated high-bandwidth interfaces necessary. Next, considering $A$ (which is expected to be done infrequently), the CIMA is loaded one 768-b row segment at a time, requiring $C_{LOAD}$=20 cycles and 768 loads in total. The cycles required for DMA transfer of each 768-b row segment is $C_A$=24 (>$C_{LOAD}$), implying 768×$C_A$=18k max. cycles for loading $A$.

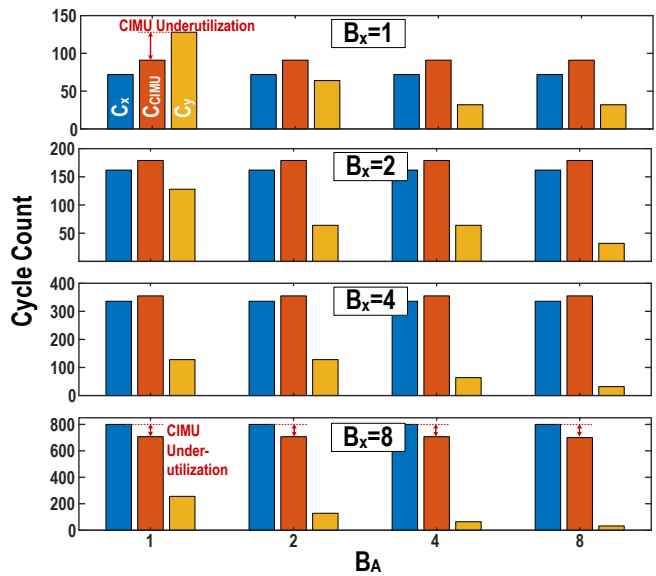

**Fig. 8. Data bandwidth analysis to and from CIMU.**

## 4. PROTOTYPE AND MEASUREMENTS

The microprocessor is implemented in 65nm CMOS (die photo in Fig. 9). Fig. 10 shows measurements of the CIMU compute. The CIMA-column transfer functions (top) are obtained by setting matrix-element bits to '1' and sweeping the number input-element bits set to '1'. For the ADC, the digitized output is then plotted, and for the ABN, the DAC analog reference that causes output-comparator transition is plotted, both showing high linearity and low variation (error bars show σ over the 256 columns). Multi-bit compute results (bottom) with uniformly-distributed input-vector and matrix elements, show excellent match with expected bit-true values and expected SQNR (from Fig. 7). Fig. 11 shows a neural-network demonstration for 1-b and 4-b input-activations/weights (topologies shown), on CIFAR-10 dataset. The chip achieves SW-simulated accuracy at energy and throughput of 5.31/105.2 μJ/class. and 176/23 images/s for 1/4-b input-activation and weight precisions. Summary and comparison tables (vs. recent neural-network accelerators) are shown.

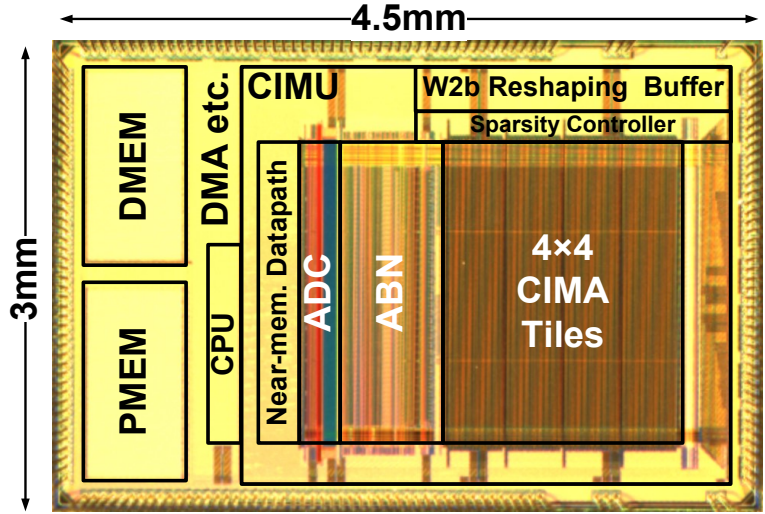

Fig. 9. Processor die photo, implemented in 65nm CMOS.

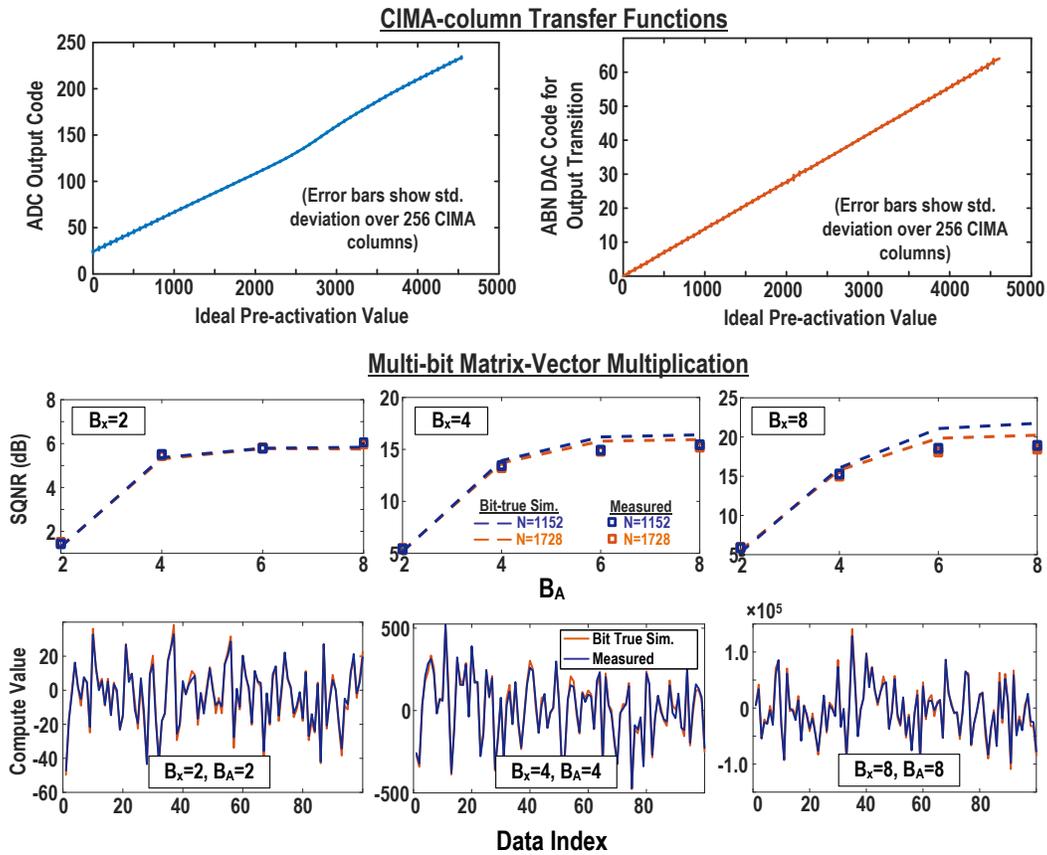

Fig. 10: Processor measurement summary.

| Summary | | | |
|---|---|---|---|
| Tech. (nm) | 65 | $F_{CLK}$ (MHz) | 100 \| 40 |
| $V_{DD}$ (V) | 1.2 \| 0.7/0.85 | Total area (mm$^2$) | 13.5 |
| Energy Breakdown @$V_{DD}$ = 1.2V \| 0.7V (P/DMEM, Reshap. Buf) , 0.85 V (rest) | | | |
| CPU (pJ/instru) | 52 \| 26 | CIMA[1] (pJ/column) | 20.4 \| 9.7 |
| P/DMEM (pJ/32b-access) | 96 \| 33 | ADC[1] (pJ/column) | 3.56 \| 1.79 |
| DMA (pJ/32b-transfer) | 13.5 \| 7.0 | ABN[1] (pJ/column) | 9.78 \| 4.92 |
| Reshap. Buf.[1] (pJ/32b-input) | 35 \| 12 | Dig. Datapath[1] (pJ/output) | 14.7 \| 8.3 |

[1]Breakdown within CIMU Acceelerator

| Neural-Network Demonstrations | | |
|---|---|---|
| | Network A (4/4-b activations/weights) | Network B (1/1-b activations/weights) |
| Accuracy of chip (vs. ideal) | 92.4% (vs. 92.7%) | 89.3% (vs. 89.8%) |
| Energy/10-way Class.[1] | 105.2 µJ | 5.31 µJ |
| Throughput[1] | 23 images/sec. | 176 images/sec. |
| Neural Network Topology | L1: 128 CONV3 – Batch norm<br>L2: 128 CONV3 – POOL – Batch norm.<br>L3: 256 CONV3 – Batch. norm<br>L4: 256 CONV3 – POOL – Batch norm.<br>L5: 256 CONV3 – Batch norm.<br>L6: 256 CONV3 – POOL – Batch norm.<br>L7-8: 1024 FC – Batch norm.<br>L9: 10 FC – Batch norm. | L1: 128 CONV3 – Batch Norm.<br>L2: 128 CONV3 – POOL – Batch Norm.<br>L3: 256 CONV3 – Batch. Norm.<br>L4: 256 CONV3 – POOL – Batch Norm.<br>L5: 256 CONV3 – Batch Norm.<br>L6: 256 CONV3 – POOL – Batch Norm.<br>L7-8: 1024 FC – Batch norm.<br>L9: 10 FC – Batch norm. |

[1]At $V_{DD}$ = 0.7V (P/DMEM, Reshap. Buf) , 0.85V (rest)

| Comparison Table | | | | | | | | | |
|---|---|---|---|---|---|---|---|---|---|
| | Not In-memory Computing | | | | In-memory Computing | | | | |
| | [3] Chen, ISSCC'16 | [4] Moons, ISSCC'17 | [5] Ando, VLSI'17 | [6] Bank., ISSCC'18 | [7] Khwa, ISSCC'18 | [8] Gon., ISSCC'18 | [9] Jiang, VLSI'18 | [1] Valavi, VLSI'18 | This work |
| Technology | 65nm | 28nm | 65nm | 28nm | 65nm | 65nm | 65nm | 65nm | 65nm |
| Area (mm$^2$) | 16 | 1.87 | 12 | 6 | unknown | 0.81 | 0.11 | 12.6 | 8.56 |
| $V_{DD}$ (V) | 0.8-1.2 | 1.0 | 0.55-1.0 | 0.8 \| 0.6 | 1.0 | 1.2 | 1.0 | 0.7,1.2 | 1.2 \| 0.85 |
| On-chip mem. | 108 kB | 128 kB | 100 kB | 328 kB | 1 kB[1] | 16 kB[1] | 2 kB[1] | 295 kB[1] | 74 kB[1] |
| Bit precision | 16 b | 4-16 b | 1-1.5 b | 1 b | 1 b | 8 b | 1 b | 1 b | 1-8 b |
| Throughput (GOPS) | 120 | 400 | 1264 | 400 \| 60 | | 8.2 | 60 | 18,876 | 4720 \| 1900[2] |
| Energy eff. (TOPS/W) | 0.0096 | 10 @4b | 6 | 535 \| 772 | 55.8 | 3.125 | 140 | 866 | 152 \| 297[2] |
| Configurability (dimensionalities, bits) | Dim. | Dim./bits | Bits | -- | -- | -- | -- | Dim. | Dim./bits |

In-memory columns [7]–[9]: **Current-domain (limited scale, configurability, accuracy)**
[1] Valavi: **Charge-domain (limited configurability)**

[1]Amount is for in-memory computing only
[2]Given for 1-b compute; scales with number of bits in input-vector elements plus matrix elements

Fig. 11. Measurement summary and comparison with prior work.

**Acknowledgements:**
This work is funded in part by Analog Devices Inc. (ADI). The authors thank E. Nestler, J. Yedida, M. Tikekar, P. Nadeau (ADI), for their extremely valuable insights and discussions.